\documentclass[journal]{IEEEtran}

\usepackage{epsfig} 
\usepackage[scaled=0.92]{helvet}
\usepackage{courier}
\usepackage{amsmath}
\usepackage{textcomp}
\usepackage{mathrsfs}
\usepackage{amssymb}
\usepackage{graphicx}
\usepackage{mathtools}
\usepackage{comment}
\usepackage{bbm}
\usepackage{url}
\usepackage{anyfontsize}
\usepackage[style=ieee]{biblatex}
\begin{document}

\title{AUTOBargeSim: MATLAB\textsuperscript{\small\textregistered} toolbox for the design and analysis of the guidance and control system for autonomous inland vessels}

\author{Abhishek Dhyani*$^{,a}$, Amirreza Haqshenas Mojaveri**, Chengqian Zhang$\dag$, Dhanika Mahipala$\ddag$, Hoang Anh Tran$\ddag$, Yan-Yun Zhang**, Zhongbi Luo**, Vasso Reppa*\\
*Delft University of Technology, The Netherlands \\ **Katholieke Universiteit Leuven, Belgium\\ $\dag$ Chalmers University of Technology, Sweden\\ $\ddag$Norwegian University of Science and Technology, Norway\\ \thanks{
 $^{a}$Corresponding author. E-mail: a.dhyani-1@tudelft.nl\\
 \copyright 2025 the authors. This work has been accepted to IFAC for publication under a Creative Commons Licence CC-BY-NC-ND. \\
 All authors contributed equally.\\
 The research leading to these results has received funding from the European Union’s Horizon 2020 research and innovation programme under the Marie Skłodowska-Curie grant agreement No 955.768 (MSCA-ETN AUTOBarge). This publication reflects only the authors’ view, exempting the European Union from any liability. Project website:http://etn-autobarge.eu/. } }

\maketitle
\begin{abstract}
This paper introduces AUTOBargeSim, a simulation toolbox for autonomous inland vessel guidance and control system design. AUTOBargeSim is developed using MATLAB and provides an easy-to-use introduction to various aspects of autonomous inland navigation, including mapping, modelling, control design, and collision avoidance, through examples and extensively documented code. Applying modular design principles in the simulator structure allows it to be easily modified according to the user's requirements. Furthermore, a GUI interface facilitates a simple and quick execution. Key performance indices for evaluating the performance of the controller and collision avoidance method in confined spaces are also provided. The current version of AUTOBargeSim attempts to improve reproducibility in the design and simulation of marine systems while serving as a foundation for simulating and evaluating vessel behaviour considering operational, system, and environmental constraints.
\end{abstract}

\begin{IEEEkeywords}
Guidance, navigation and control (GNC) of marine vessels; Nonlinear and optimal control in marine systems; Modeling, identification, simulation, and control of marine systems
\end{IEEEkeywords}

\section{Introduction}

Marine vessels play an undeniably important role in freight transportation, accounting for more than two-thirds of the modal share in the European Union \cite{Eurostat_2024}. Within the maritime domain, the research and development of autonomous vessel technologies have received increased interest due to their potential to improve safety and energy efficiency. Inland waterway transportation connects many of the EU's major cities and industrial regions by rivers and canals, spanning more than 31,000 kilometres. Improving the autonomy of inland waterway vessels (IWVs) offers a unique opportunity to contribute to improving the safety of the inland waterway transportation network, which involves navigating within complex environments and narrow waterway boundaries.

The guidance, navigation, and control (GNC) system plays a crucial role in the safe and reliable operation of ASVs. This system comprises technologies ranging from modern sensors to complex algorithms and software that enable the ASV to perceive its environment and have situational awareness and decision-making capabilities. Furthermore, simulation-based testing of the GNC system is an essential step in the design process. A few freely available scientific simulation software/ toolboxes have been proposed for maritime simulation (See e.g., \cite{perez2006overview,sukas2019theoretical,blindheim2021electronic,krasowski2022commonocean,tengesdal2023simulation,CLEMENT2024147}). Arguably, the marine systems simulator (MSS) \cite{perez2006overview} is the most popular and widely used MATLAB\textsuperscript{\small\textregistered} -based toolbox, consisting of various classes of models, transformation functions, guidance and control algorithms, among others. More recently, simulation toolboxes focusing on evaluating collision avoidance algorithms have also been proposed (see \cite{krasowski2022commonocean,tengesdal2023simulation,CLEMENT2024147}). 

While the existing simulation platforms are well-equipped with many of the necessary functionalities for autonomous vessel simulation, the majority of these platforms consider open-sea simulation only. However, confined waterways such as inland waterways, ports and canals, which are key use cases for the deployment of autonomous vessels, are not considered. Operating vessels in confined waters is particularly challenging as they are constrained by several factors, such as canal width, infrastructures, dynamic water levels, river currents, and riverbed variations. The existing platforms rely on mathematical models to simulate vessel maneuvers that mimic the characteristics of a seagoing vessel. The hydrodynamic forces generated due to shallow water depth in inland waterways can significantly impact the vessel’s motion and maneuverability. By default, these platforms do not offer quantitative performance indicators for the evaluation of guidance and control algorithms. These metrics provide useful benchmarking data for comparing various algorithms to state-of-the-art methods.

In this work, we address these gaps by introducing AUTOBargeSim, a MATLAB\textsuperscript{\small\textregistered}-based simulation toolbox for the design and evaluation of guidance and control algorithms for autonomous inland navigation. AUTOBargeSim has been created with a focus on modularity, reproducibility, and ease of use as the key design principles and is freely available for research and educational purposes. It allows the users to visualize inland map features, set up scenarios for vessel navigation, select various parameters, simulate the vessel motion, and evaluate the performance of vessel path following and collision avoidance algorithms. Various examples provide an introduction to applying specific classes and methods from the toolbox based on the user's requirements.

The remainder of the paper is organised as follows: in Section 2, instructions for installing and using the simulator toolbox are given. Its design and structure as well as the comprising methods and parameters, are detailed in Section 3. Further, qualitative metrics for performance evaluation are presented and described in Section 4. Finally, the conclusions and scope for future development are discussed in Section 5.

\section{Installation and Usage}
AUTOBargeSim can be downloaded from its GitHub repository\footnote{https://github.com/AUTOBarge/autobargesim}. After extracting the .zip file to your desired directory, the toolbox can be installed by simply running the file $\texttt{install\_absim.m}$ and following the instructions in the MATLAB\textsuperscript{\small\textregistered} command window. MATLAB\textsuperscript{\small\textregistered}  Control System Toolbox{\small\texttrademark}, Mapping Toolbox{\small\texttrademark}, and Statistics and Machine Learning Toolbox{\small\texttrademark} are prerequisites for using AUTOBargeSim. Furthermore, the CasADi toolkit \cite{Andersson2019} for automatic differentiation is used at the backend for implementing optimal control problems within the framework of Model Predictive Control (MPC). 

\begin{figure}
    \centering
    \includegraphics[width=0.7\linewidth]{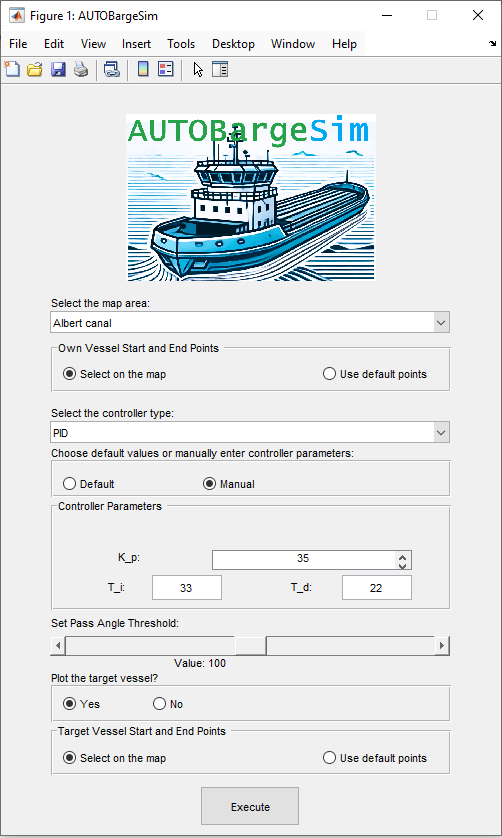}
    \caption{AUTOBargeSim Graphical User Interface}
    \label{fig:gui}
\end{figure}

The toolbox includes several tutorials for easily customizing the methods included in the provided modules. These tutorials are in the respective module directories, as summarized in Table \ref{tab:Tutorials}. Furthermore, the GUI (as shown in Fig. \ref{fig:gui}) allows a simple and fast execution of the toolbox, with the possibility to adjust some critical inputs, such as the map area, controller gains, etc. It can be executed by running the script \texttt{main\_gui.m}, selecting between various options in the GUI window and pressing the \textit{Execute} button. It should be noted that the GUI currently has a limited number of inputs; however, the underlying methods allow far more input flexibility. 

\begin{table}
    \centering
    \caption{List of tutorial files for the respective modules}
    \begin{tabular}{|c|c|c|} \hline 
         Module&  File name\\ \hline 
         \texttt{maps}&  \texttt{maps\_test\_demo.m}  \\ \hline 
         \texttt{control}&  \texttt{control\_demo\_PID.m} \\ \cline{2-2} 
         &  \texttt{control\_demo\_MPC.m} \\ \cline{2-2} 
         &  \texttt{control\_demo\_PID\_guidance.m} \\ \cline{2-2} 
 & \texttt{control\_demo\_MPC\_guidance.m}\\\hline
 \texttt{model\&actuator}  & \texttt{demo1.m}\\ \cline{2-2}
 & \texttt{demo2.m}\\\cline{2-2}
 & \texttt{demo3.m}\\\hline
 \texttt{guidance}& \texttt{demo.m}\\\hline
 \texttt{colav}& \texttt{demo.m}\\\hline
    \end{tabular}
    \label{tab:Tutorials}
\end{table}

\section{Design and Structure}

The simulator follows a modular design. Based on functionality, it is divided into several modules, and under each module, various algorithms and demos are provided.
\begin{figure*}
    \centering
    \includegraphics[width=0.6\linewidth]{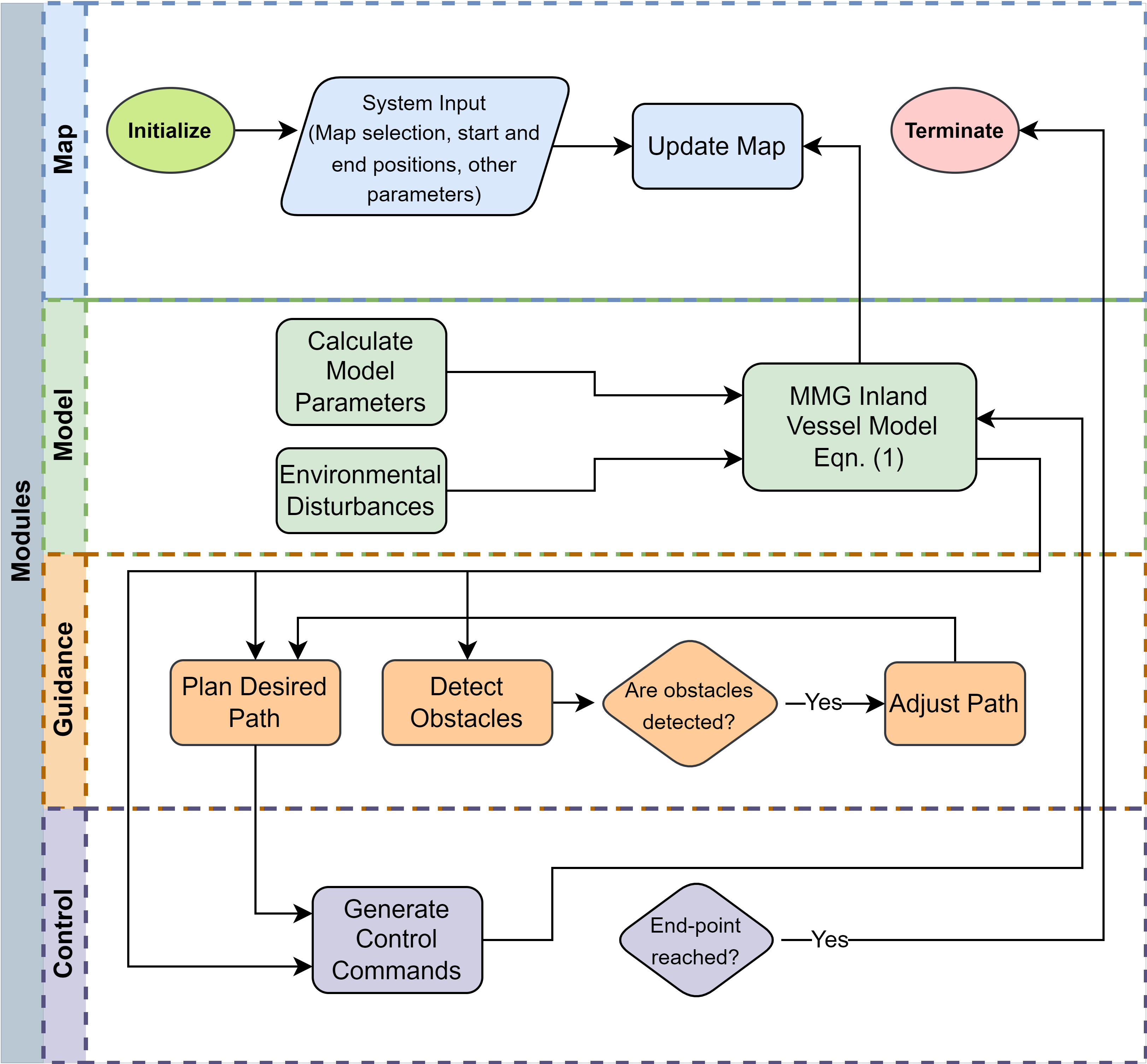}
    \caption{Flowchart depicting the various modules and the flow of control}
    \label{fig:flowchart}
\end{figure*}

The overall structure and the connection between various modules are illustrated in Fig. \ref{fig:flowchart}.
The mathematical model describing the IWV dynamics and the actuators is defined in the Model module while considering environmental disturbances, including currents and shallow-water effects. Static environmental data, including waterway boundaries and the locations of static objects along the path, is extracted from chart files within the Map module. The map module also provides a set of initial waypoints and, subsequently, a desired path between two selected points on the map, which serve as the inputs to the Guidance module. The Guidance module employs a guidance law to compute the desired/ reference course angles for vessel navigation. Next, the reference course angles are provided to the Control module, and a control law computes the desired rudder angle for steering the vessel \cite{Zhang2024}. In the case of a collision avoidance scenario, the motion of a \textit{target vessel} is also simulated, and the \textit{own vessel} performs a collision avoidance maneuver if required. The simulation terminates when the vessel successfully reaches the provided endpoint. Finally, all the processed data is used to update the map visualization, and the various performance metrics are displayed. Each module is explained in detail in the subsections that follow.

\subsection{Model Module}
Inland vessels frequently operate on confined waterways in the presence of dynamic traffic and hydraulic structures such as bridges and locks. Therefore, a reliable maneuvering model for accurately predicting the vessels' dynamics is critical for safe navigation. 

\begin{figure}
    \centering
    \includegraphics[width=1\linewidth]{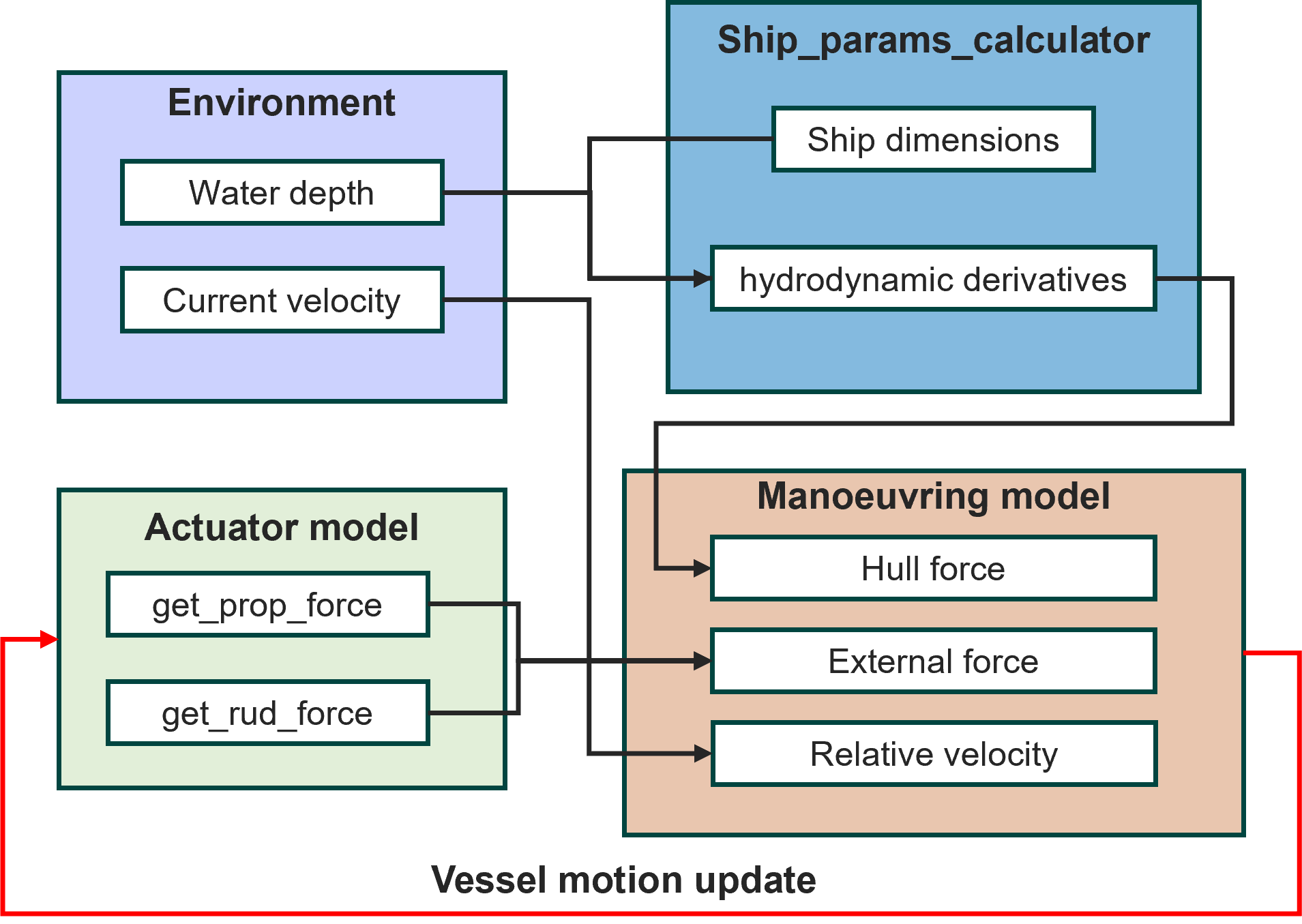}
    \caption{Architecture of manoeuvring and actuator module.}
    \label{fig:manoeuvring}
\end{figure}

The maneuvering model follows the popularly known Manoeuvring Modelling Group (MMG) model \cite{ogawa1978mathematical} architecture, where the hydrodynamic forces and moments are derived into individual components. The original MMG model was developed for classic commercial vessels in open water. Due to its flexible and modular structure, the model can be extended by incorporating shallow water effects to account for the influencing factors of inland waterways.
The rigid body dynamics can be represented as: 
\begin{equation}
\begin{aligned}
    (m + m_x) \dot{u} - (m + m_y) v r - x_G m r^2 &= X_H + X_P + X_R,  \\
    (m + m_y) \dot{v} - (m + m_x) u r + x_G m \dot{r} &= Y_H + Y_R,  \\
    \left( I_z + x_G^2 m + J_z \right) \dot{r} + x_G m (\dot{v}+ur) &= N_H + N_R, 
\end{aligned}
\end{equation}
where the left side contains the mass $(m,m_x,m_y)$ and inertia terms $(I_z,J_z)$; $({u},{v},{r})$ represent the surge, sway velocity and the yaw rate; $(X,Y,N)$ denote the summation of the surge force, the sway force, and the yaw moment. 
The subscripts $(H,P,R)$ represent the hydrodynamic force of the individual components acting on the hull, propeller, and rudder, respectively. Therefore, the model is divided into two classes: \texttt{modelClass.m} calculates the hydrodynamic forces acting on the hull, and \texttt{actuatorClass.m} calculates the propeller thrust and rudder forces (see Fig. \ref{fig:manoeuvring}).

It should be noted that the shallow water effect is included by the modified terms acting on the ship hull  $(X_H,Y_H,N_H)$, including the added resistance and sway force due to the reduced under-keel clearance. 
\subsubsection{Hydrodynamic Forces on the Hull}
\hfill \break
The force and moment acting on the hull can be calculated as:
\begin{equation}
\begin{aligned}
    X_H &= 0.5 \rho L T U^2 X_H', \\
    Y_H &= 0.5 \rho L T U^2 Y_H', \\
    N_H &= 0.5 \rho L^2 T U^2 N_H',
\end{aligned}
\end{equation}
where $\rho$ is the fresh water density, $L$ is the vessel length, $T$ is the vessel draught, $U$ is the total speed, and $(X_H',Y_H',N_H')$ are non-dimensional forces and moment, given as:
\begin{align}
    X_H' &= -R_0' \cos^2 \beta_m + X_{\beta \beta}' \beta_m^2 + X_{\beta r}' \beta_m r' 
    + X_r' r'^2 \notag \\
    &\quad + X_{\beta \beta \beta \beta}' \beta_m^4, \\
    Y_H' &= Y_{\beta}' \beta_m + Y_r' r' + Y_{\beta \beta \beta}' \beta_m^3 
    + Y_{\beta \beta r}' \beta_m^2 r' \notag \\
    &\quad + Y_{\beta r r}' \beta_m r'^2 + Y_{r r r}' r'^3, \\
    N_H' &= N_{\beta}' \beta_m + N_r' r' + N_{\beta \beta \beta}' \beta_m^3 
    + N_{\beta \beta r}' \beta_m^2 r' \notag \\
    &\quad + N_{\beta r r}' \beta_m r'^2 + N_{r r r}' r'^3,
\end{align}
where $-R_0'$ is the resistance coefficient including shallow water effect \cite{zhang2023development}, and $(X_{\beta \beta}',...,N_{r r r}')$ are the so called hydrodynamic derivatives which can be calculated from the function \texttt{ship\_params\_calculator} using empirical formulas based on ship dimensions.

\subsubsection{Propeller and Rudder Forces}
\hfill \break
The actuator module (\texttt{actuatorClass}) is based on a conventional propeller-rudder configuration. The thrust of a ducted propeller can be calculated using the function (\texttt{get\_prop\_force}) based on the equation:
\begin{equation}
X_P = (1-t) \rho n_P^2D_P^4K_T(J), \\
\end{equation}
where $t$ is the thrust deduction, $n_P$ is the propeller rpm, $D_P$ is the propeller diameter and $K_T(J)$ is the thrust coefficient as a function of advanced ratio $J$. In this work, the open water coefficient is derived from the open-source propeller design tool OpenProp \cite{epps2009openprop}.

The rudder steering force and moment are calculated using function (\texttt{get\_rud\_force}) as follows:

\begin{equation}
\begin{aligned}
    X_R &= -(1 - t_R) (F_{N}^P + F_{N}^S) \sin \delta, \\
    Y_R &= -(1 + \alpha_H) (F_{N}^P + F_{N}^S) \cos \delta, \\
    N_R &= - (x_R + \alpha_H x_H) (F_{N}^P + F_{N}^S) \cos \delta,
\end{aligned}
\end{equation}
where $F_N$ is the rudder normal force, the superscript $P$ and $S$ denote the rudder at port side and starboard side, $t_R$ denotes the steering resistance deduction factor, $\alpha_H$ represents the rudder force increase factor, $x_R$ is the relative location of rudders and keeping identical at each side, $x_H$ is the relative acting point of the additional lateral force, and $\delta$ is the rudder angle. Here, $F_N$ is given as:
\begin{equation}
F_N=0.5\rho A_RU_R^2\left(\frac{6.13\mathrm{\Lambda}}{\mathrm{\Lambda}+2.25}\sin \alpha_R \right),
\end{equation}
where $A_R$ is the rudder area, $U_R$ is the incoming flow velocity at the rudder ($U_R=\sqrt{u_R^2+v_R^2}$), $\mathrm{\Lambda}$ is the rudder aspect ratio and $\alpha_R$ is the effective inflow angle at the rudder during manoeuvring (see \cite{ogawa1978mathematical}).

\subsection{Map Module}
\begin{figure}
    \centering
    \includegraphics[width=1\linewidth]{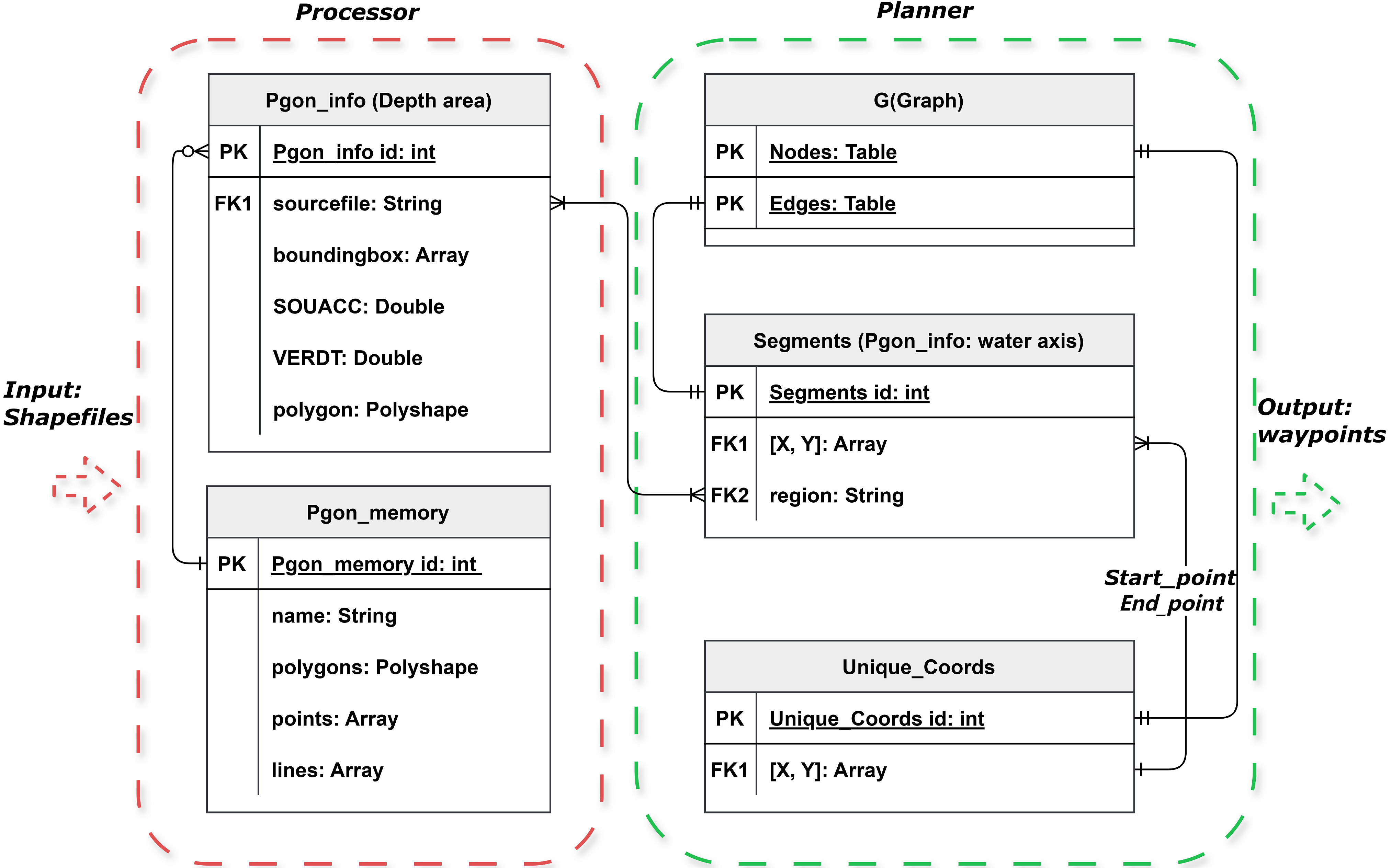}
    \caption{Entity Relationship Diagram (ERD) shows the relationship between different data in each submodule of the map module.}
    \label{fig:erd}
\end{figure}

\begin{figure}
    \centering
    \includegraphics[width=1\linewidth]{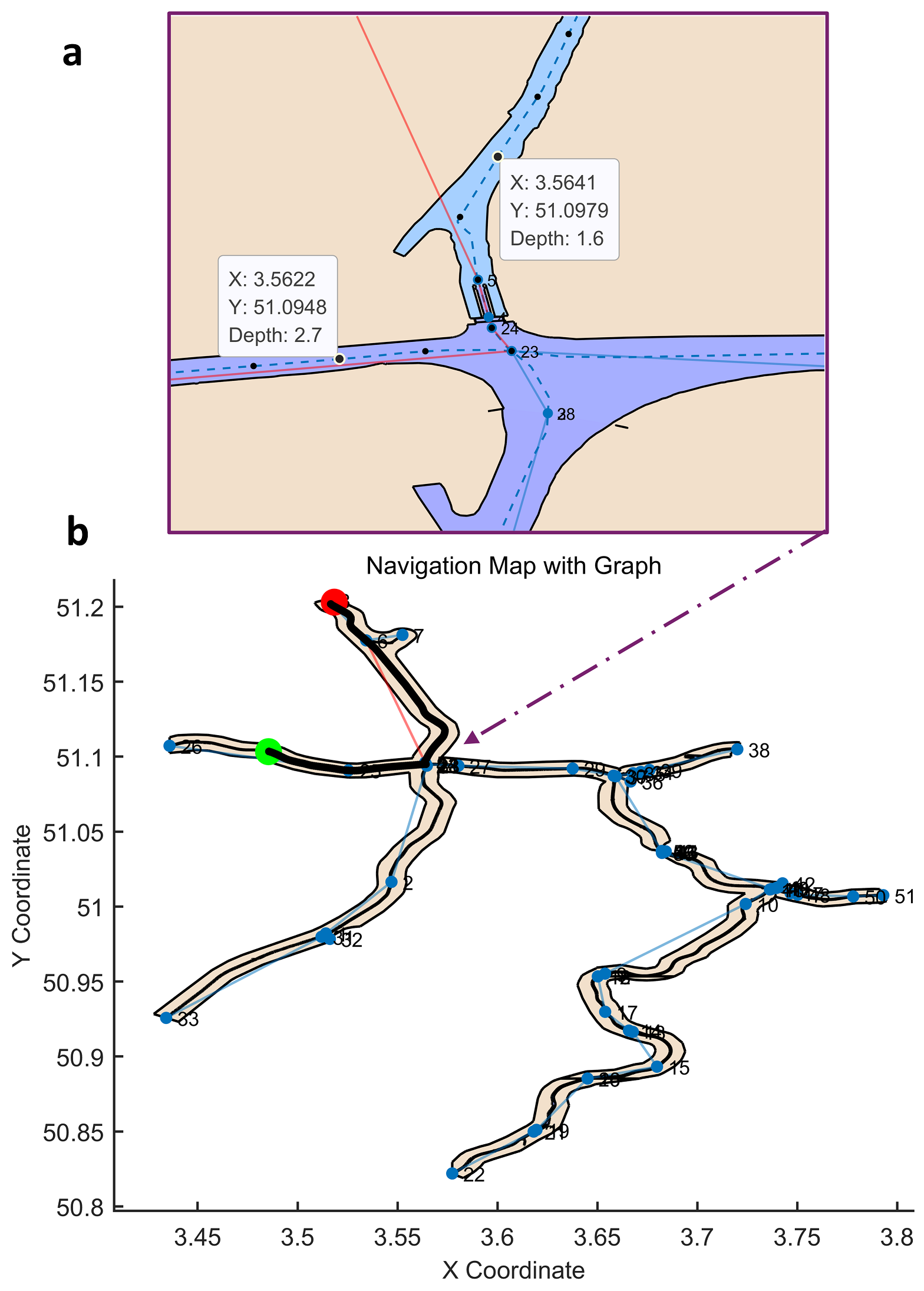}
    \caption{(a) A localized, zoomed-in view of the navigation map, displaying the coordinates and depth information of waypoints within the waterway. (b) The navigation map for the Ghent area includes the starting point (green), end point (red), the waypoints (black) and the graph composed of nodes and edges.}
    \label{fig:map}
\end{figure}

The Map module processes Inland Electronic Navigational Chart (IENC) data to provide essential environmental information for path planning and collision avoidance. It comprises two main submodules: the Processor and the Planner. The Processor converts input shapefile (\texttt{.shp}) data into a structured format called \texttt{pgon\_memory}, while the Planner uses this data to generate waypoints with associated depth information, as shown in Fig. \ref{fig:map}a. Fig. \ref{fig:erd} illustrates the relationships between data entities in these modules.

Due to MATLAB's inability to decode S-57 standard ENC files (\texttt{.000} files), which contain standardized navigational data with semantic tags and attributes, we employ a preprocessing step using GDAL \cite{GDAL2020} Python library. This step converts \texttt{.000} files into shapefiles, enabling the Processor to manage the data efficiently. By extracting specified regions and attributes from the IENC and structuring them into spatial data, compatibility with the simulator is ensured.

\subsubsection{Processor}
\hfill \break
The Processor reads shapefiles generated from the \texttt{.000} ENC files and categorizes them based on predefined navigational features. Users are allowed to add predefined features from the S-57 standard as needed. The following are the main features: Depth Areas (\texttt{depare}): Polygons with attributes such as Sounding Accuracy (\texttt{SOUACC}), Vertical Datum (\texttt{VERDAT}), and polygon extent (\texttt{boundingbox}), providing essential depth and positional data. 
Waterway Axes (\texttt{wtwaxs}): Polylines indicating central navigational paths within waterways.
Bridges (\texttt{bridge}): Polygons representing bridge locations and dimensions, important for height restrictions.
Land Areas (\texttt{lndare}): Polygons denoting non-navigable regions.

For each category, the Processor reads geometric data and attributes from the shapefiles, constructing the \texttt{pgon\_memory} structure with fields: \texttt{name}: Category name (e.g., \texttt{depare}, \texttt{wtwaxs}).
 \texttt{points}: Coordinates of point features. \texttt{lines}: Coordinate arrays of line features. \texttt{polygons}: \texttt{polyshape} objects of polygon features. \texttt{info}: Attributes for each geometric entity.

For depth areas, the Processor constructs \texttt{polyshape} objects and extracts depth attributes like \texttt{SOUACC} and \texttt{VERDAT}, as well as the \texttt{boundingbox} and the unique identifier of the source \texttt{.000} file (\texttt{sourcefile} or \texttt{region}), storing them in \texttt{depare.info}. For processing waterway axes, it extracts coordinates, removes redundant points to maintain data integrity, and splits the data into multiple segments, associating each line with relevant metadata (\texttt{region}).

\subsubsection{Planner}
\hfill \break
The Planner uses \texttt{pgon\_memory} to perform path planning by constructing a navigational graph denoted as $G$ illustrated in Fig. \ref{fig:map}b. This graph comprises nodes and edges derived from \texttt{wtwaxs}. The nodes are represented by unique coordinates stored in \texttt{unique\_coords}, which are extracted from the segment startpoints and endpoints of the waterway axes. Edges between nodes are established based on the connectivity of the segments, forming a comprehensive representation of the navigable waterways.

To ensure the graph $G$ is fully connected, the Planner checks for disconnected components using functions \texttt{conncomp} and connects them by identifying the closest nodes between components. Depth information from the depth areas (\texttt{depare}) is associated with the nodes and edges of the graph through spatial queries. Specifically, the Planner uses functions \texttt{isinterior} to perform a hierarchical search from \texttt{region} to \texttt{boundingbox} to \texttt{polygon} to determine if nodes or path points lie within depth polygons and assigns depth values from attributes \texttt{SOUACC} and \texttt{VERDAT} to the corresponding nodes and edges.

Path planning is performed by finding the shortest path in the graph $G$ that satisfies depth constraints using the Dijkstra algorithm \cite{10.1145/3544585.3544600}. The Planner considers the given starting point (\texttt{given\_point1}) and ending point (\texttt{given\_point2}), locating the nearest nodes in \texttt{unique\_coords} and planning a path between them. The resulting path is stored in \texttt{path\_points}, with the corresponding depth information saved as \texttt{path\_depths}.

To refine the path for practical navigation, the Planner removes duplicate points and smooths the trajectory. Functions \texttt{removeDuplicatePoints} and \texttt{smoothPoints} are used to remove redundant points and provide a feasible route for the vessel. The final output is a series of waypoints in \texttt{path\_points} with corresponding depth information in \texttt{path\_depths}, ensuring the planned path is safe and efficient given the vessel's draft and environmental constraints.
\subsection{Guidance Module}
The main goal of the Guidance module is to fulfil two tasks: Track Keeping and Collision Avoidance. The Track Keeping submodule ensures that the ship sails toward the next waypoint. The Collision Avoidance submodule adjusts the speed and course angles provided by the Track Keeping submodule of the ship to avoid obstacles.

\subsubsection{Track Keeping}
\hfill \break
The Track Keeping submodule receives the waypoint list from the map class and provides a course angle and speed reference for Collision Avoidance to ensure that the ship follows the desired waypoints.

In this submodule, we define the \texttt{track-keeping} class as a superclass containing a common function that can be used by subclasses created for specific track-keeping controllers. In this class, we provide a function to find the current active waypoint in the waypoint list and the position of the ship. The key properties are the radius of acceptance, $R_a$, and \texttt{pass\_angle\_threshold}, which are used to identify if the ship passed a waypoint.

The default track-keeping controller is the Line-of-Sight (LOS) steering algorithm \cite{fossen11}. The only property of this subclass is the proportional gain of the LOS steering law $K_{p_{los}}= 1/D_{los}$, with $D_{los}$ being the look-ahead distance. The main function is \texttt{compute\_LOSRef}, which receives the current state of the vessel, a \texttt{waypoint\_list} containing the coordinates of the waypoints, and the expected nominal speeds at each waypoint. It then returns the reference course angle, $\chi_d$.

\subsubsection{Collision Avoidance}
\hfill \break
The collision avoidance submodule is responsible for modifying the course and speed commands provided by the Track Keeping submodule to avoid collisions. This submodule contains two main classes: a superclass named \texttt{colav} and a subclass named \texttt{sbmpc}. The \texttt{colav} class includes common methods necessary for implementing any collision avoidance algorithm, such as internal kinematic models for trajectory prediction. Users are encouraged to use these methods to implement custom collision avoidance algorithms in this simulator. End-users interact with the application layer of the module, which includes classes related to specific collision avoidance algorithms. Currently, only one such class is available in this simulator: the class implementing the Scenario-based Model Predictive Control (SBMPC) algorithm \cite{johansen_ship_2016}. The SB-MPC algorithm presents a proactive collision avoidance strategy based on simulation and receding horizon optimization. It guarantees compliance with COLREGS Rules 6, 8, and 13-19 and mitigates collision risks by evaluating a cost function that accounts for the predicted trajectories of both the ship and obstacles. The algorithm's cost function is comprehensive, factoring in COLREGS rule violations, maneuvering effort, and collision risk. Solutions are generated from a finite set of discrete options through an exhaustive search method. The exact details and components of the algorithm are provided in \cite{johansen_ship_2016, hagen_mpc-based_2018, mahipala_model_2023}.

The \texttt{sbmpc} class contains the methods and properties related to the use of the SBMPC algorithm. Its key properties include \texttt{T} and \texttt{dt}, which represent the prediction time horizon and the sample time of the algorithm, respectively. Apart from these mandatory inputs, properties such as \texttt{T\_chi} and \texttt{T\_U} are optional arguments when initializing an \texttt{sbmpc} class object. They represent the course and speed time constants of the internal vessel kinematic model, which is adapted from \cite{ship_colav_and_anti_grounding}. Additionally, a property named \texttt{tuning\_param} contains the constant tuning parameters listed in \cite{johansen_ship_2016} as sub-properties, which can be modified using optional arguments when initializing the \texttt{sbmpc} object. The class only has one public method that is accessible for the user, which is named \texttt{run\_sbmpc()}. The function takes the current state of the vessel, course, and speed commands from the Track Keeping submodule, course and speed modifications from the previous time step, and the states of one or more target vessels as input. The outputs of the function are course and speed modifications for the current time step.


\subsection{Control Module}
The Control module provides two low-level controllers, namely Proportional-Integral-Derivative (PID) control and Model Predictive Control (MPC), to calculate the required rudder angle for navigating the vessel and ensuring that it tracks the desired heading angle. The Control module is defined as a class containing several properties and methods. The properties include \texttt{num\_st}, \texttt{num\_ct}, and \texttt{Flag\_cont}, which represent the number of state variables, the number of control variables, and the selected control method, respectively. It also includes \texttt{pid\_params}, a structure data type containing the PID controller's parameters including $K_p$: proportional gain, $T_i$: integral time-constant, $T_d$: derivative time-constant, and \texttt{psi\_d\_old}, \texttt{error\_old} for storing the desired heading angle and heading angle error from the last iteration. Finally, the structure \texttt{mpc\_params} contains the MPC controller's parameters including $T_s$: sampling time, $N$: prediction horizon, \texttt{headingGain}: weighting gain for heading angle, \texttt{rudderGain}: weighting gain for rudder angle, \texttt{max\_iter}: maximum number of iterations for the MPC and \texttt{deltaMAX}: maximum value for rudder angle. The methods within this class manage tasks such as handling the state variables and updating the control parameters. The property \texttt{states} represents the controlled state variables of the system, $s=\left[r, \psi \right]^{T}$. Further, the variables $\psi_d$ and $r_d$ stand for the desired heading angle and turning rate, which are the outputs of the Guidance module, and form the reference input vector, $s_{\text{ref}}=\left[r_d, \psi_{d} \right]^{T}$. A detailed explanation follows in the respective subsections below.
\subsubsection{PID Controller}
\hfill \break
The PID controller is a classical control technique popular for its simplicity and ease of implementation. Within the Control module, the PID controller is the default controller choice. It determines the rudder angle by minimizing the error between the desired and actual heading angles. The control law can be stated as:
\begin{equation}
    \delta_c (t) = K_p \psi_e(t)+T_d(\psi_e(t)-\psi_e(t-1))+ \frac{1}{T_i}\left( \sum_{i=0}^{t}\psi_{e_i}\right),
\end{equation}
where $\psi_e(t)=\psi(t)-\psi_{d}(t)$ is the heading angle error, and $K_p$, $T_i$ and $T_d$ are the controller's proportional gain, and the integral and derivative time constants, respectively. The method 
\texttt{LowLevelPIDCtrl} computes and outputs the desired rudder command $\delta_c$ by using the $\psi$ variable from \texttt{states} and the reference heading angle $\psi_d$ as inputs, respectively.
\subsubsection{Model Predictive Control (MPC)}
\hfill \break
This subsection describes the implementation of MPC within the Control module. The MPC determines the rudder control input for the vessel based on the desired heading angle and turning rate. The Control module includes multiple methods for formulating the MPC for rapid implementation. These methods include \texttt{init\_mpc}, \texttt{initial\_guess\_creator}, \texttt{constraintcreator} and \texttt{LowLevelMPCCtrl}. The \texttt{init\_mpc} method employs CasADi as the backbone to formulate a graph stored in \texttt{mpc\_nlp} 
for solving the constrained Nonlinear Programming (NLP) problem defined within the MPC. The \texttt{initial\_guess\_creator} method requires two inputs, the initial states and the initial control input, to construct the initial guess vectors and store them in an internal structure. The \texttt{constraintcreator} obtains its necessary information from \texttt{mpc\_params} and generates a built-in structure for storing all the needed arguments to be passed on to the NLP solver created by the \texttt{init\_mpc}. The method for building the MPC controller is \texttt{LowLevelMPCCtrl}, which uses \texttt{states}, $s_{\text{ref}}$, \texttt{args}, \texttt{initial\_guess} and \texttt{mpc\_nlp} as its inputs. The variable \texttt{args} is the output of the \texttt{constraintcreator} method and presents the NLP arguments. Note that this method is required to be called at each iteration of the simulation and solves the following optimization problem:
\begin{equation}\label{eq:mpcopt}
\begin{aligned}
&\min_{\delta_c} J(s,s_{\text{ref}},\delta_{c}, k) \\ \text{subject to} \ & s \left[ k+1 \right] = f(s[k], \delta_{c}[k]), \\ &s[0] = s_{0}, \\ & s \in U_{\mathbf{s}}, 
\end{aligned}
\end{equation}
where $f(s[k], \delta_{c}[k])$ denotes the prediction model describing the relation between the states and the input, and $U_s \subset \mathbb{R}^2$ represents the set of permissible states \cite{fossen11}. Moreover, $J$ represents the cost function and can be formulated as 
\begin{equation}
\begin{aligned}
J(s,s_{\text{ref}},\delta_{c},k) = \sum_{i=1}^{N} & \left[ (s-s_\text{ref})_{(k+i)}^T Q (s-s_\text{ref})_{(k+i)} + \right. \\ & \left. \left(\delta_{c(k+i-1)} \right)^2 p \right],
\end{aligned}
\end{equation}

where $Q \in \mathbb{R}^2$ and $p\in \mathbb{R}$ are the state- and control- weights, respectively, and they are chosen by the designer. Solving the optimization problem \eqref{eq:mpcopt} yields the optimal rudder angle for the time instance $k+1$.

\section{Performance Evaluation}
The guidance and control module includes functions used to evaluate the controllers, namely the track-keeping and path-following controllers.
\subsection{Track-keeping Control}
The \texttt{perf} function under \texttt{track-keeping} class provides the following indices:
\begin{enumerate}
    \item \textbf{Nominal distance ($D_{\text{nominal}}$)}: The cumulative distance between waypoints from the start point to the endpoint, and is computed as:
    \begin{align}
        D_{\text{nominal}} = \sum_{i=1}^{N-1}\sqrt{(x^{wp}_{i+1}-x^{wp}_{i})^2+(y^{wp}_{i+1}-y^{wp}_{i})^2},
    \end{align}
    where $[x^{wp}_{i} y^{wp}_{i}]$ is the position of $i^{th}$ waypoint, and $N$ is the number of waypoints.
    \item \textbf{Nominal navigation time ($T_{\text{nominal}}$)}: The ``unreal'' time it takes for the vessel to sail from the start point to the endpoint with the nominal speed, $v_{\text{ref}}$. The nominal navigation time is calculated as follows:
    \begin{equation}
        T_{\text{nominal}} = \frac{D_{\text{nominal}}}{v_{\text{ref}}}.
    \end{equation}
    \item \textbf{Actual navigation distance ($D_{\text{actual}}$)}: The actual navigation distance that the vessel sails from the start point to the endpoint, and is computed as:
    \begin{align}
        D_{\text{actual}} = \sum_{i=1}^{\mathcal{I}-1}\sqrt{(x_{i+1}-x_{i})^2+(y_{i+1}-y_{i})^2},
    \end{align}
    where $[x_{i} y_{i}]$ is the position of the ship at time $i$, and $\mathcal{I}$ is the total simulation iteration taken for the vessel to reach the endpoint.
    \item \textbf{Actual navigation time ($T_{\text{actual}}$)}: The ``unreal'' time that it takes for the vessel to sail from the start to the endpoint and is calculated as follows:
    \begin{equation}
        T_{\text{actual}} = \Delta T \times \mathcal{I},
    \end{equation}
    where $\Delta T$ is the sampling period of the simulation.
\end{enumerate}
\subsection{Path Following Control}
To evaluate the controller's performance for the vessel's path following control, the following key performance metrics have been utilized:
\begin{enumerate}
    \item \textbf{Cumulative Heading error} ($\psi_{e,c}$): The cumulative heading error $\psi_{e,c}$ is calculated by
    \begin{equation}
    \psi_{e,c}=\int_{t=1}^{N}(\psi(t)-\psi_{d}(t)) dt
    \end{equation}
where $\psi_{d}(t)$ is the desired heading angle at the time instant $t$.
    \item \textbf{Cumulative cross-track error} (CXTE): The Cumulative cross-track error is calculated by 
    \begin{equation}
    \text{CXTE}=\int_{t=1}^{N}\sqrt{(x(t)-x_{cl}(t))^2+(y(t)-y_{cl}(t))^2} dt,
    \end{equation}
where $(x_{cl}(t),y_{cl}(t))$ are the points of the desired path that are at the closest distance from the vessel's position at the time instant $t$.
\end{enumerate}

\section{Conclusion and Future works}
This paper presented AUTOBargeSim, a toolbox for simulating autonomous inland vessels.
AUTOBargeSim provides a vital environment for testing various aspects of autonomous vessel navigation in inland waterway environments.
Its modular design provides improved flexibility, allowing users to easily modify or replace individual modules without impacting the functionality of others. Further, AUTOBargeSim is extensively documented and openly available, promoting reproducibility in the design and development of marine navigation systems.

The future developments of the simulator will aim to incorporate additional aspects of autonomous vessel operations, such as considering sensor characteristics and abnormal events.
A communication module will be developed to allow information exchange between vessels, providing collaborative navigation capabilities.
Moreover, the collision avoidance system will be evaluated against metrics suitable for inland waterway scenarios. Currently, the toolbox supports only constant-speed vessel simulations; however, it is planned to include variable-speed maneuvering capabilities to better reflect real-world operational characteristics.
\printbibliography

\end{document}